\documentstyle[aaspp4,12pt]{article}
\begin{document}

\def\P{\bar{\Phi}}
\def\st{\sigma_{\rm T}}
\def\vk{v_{\rm K}}
\def\s{\ifmmode \sigma \else $\sigma~$\fi}
\def\ss{\ifmmode \sigma_s \else $\sigma_s~$\fi}
\def\sles{\lower2pt\hbox{$\buildrel {\scriptstyle <}
   \over {\scriptstyle\sim}$}}
\def\sgreat{\lower2pt\hbox{$\buildrel {\scriptstyle >}
   \over {\scriptstyle\sim}$}}
\def\lsim{\lower2pt\hbox{$\buildrel {\scriptstyle <}
   \over {\scriptstyle\sim}$}}
\def\gsim{\lower2pt\hbox{$\buildrel {\scriptstyle >}
   \over {\scriptstyle\sim}$}}

\title{Towards Resolving the Crab $\sigma-$Problem: 
A Linear Accelerator?}

\author{Ioannis Contopoulos \altaffilmark{1} and
Demosthenes Kazanas \altaffilmark{2}}
\affil{}
\altaffiltext{1}{200 Akti Themistokleous Str., Piraeus 18539, 
Greece; e-mail: jkontop@elval.vionet.gr}
\altaffiltext{2}{NASA/GSFC, Greenbelt, MD 20771, 
USA; e-mail: kazanas@milkyway.gsfc.nasa.gov}

\begin{abstract}

Using the exact solution of the axisymmetric pulsar magnetosphere
derived in a previous publication and the conservation laws of the
associated MHD flow, we show that the Lorentz factor of the outflowing
plasma increases linearly with distance from the light
cylinder. Therefore, the ratio of the Poynting to particle energy
flux, generically referred to as $\sigma$, decreases inversely
proportional to distance, from a large value (typically $\gsim 10^4$)
near the light cylinder to $\s \simeq 1$ at a transistion distance
$R_{\rm trans}$.  Beyond this distance the inertial effects of the
outflowing plasma become important and the magnetic field geometry
must deviate from the almost monopolar form it attains between
$R_{lc}$ and $R_{\rm trans}$.  We anticipate that this is achieved by
collimation of the poloidal field lines toward the rotation axis,
ensuring that the magnetic field pressure in the equatorial region
will fall-off faster than $1/R^2$ ($R$ being the cylindrical
radius). This leads both to a value $\s=\ss \ll 1$ at the nebular
reverse shock at distance $R_s$ ($R_s \gg R_{\rm trans}$) and to a
component of the flow perpendicular to the equatorial component, as
required by observation. The presence of the strong shock at $R = R_s$
allows for the efficient conversion of kinetic energy into
radiation. We speculate that the Crab pulsar is unique in requiring
$\ss \simeq 3 \times 10^{-3}$ because of its small translational
velocity, which allowed for the shock distance $R_s$ to grow to values
$\gg R_{\rm trans}$.
%In other pulsars, e.g. Vela,
%$R_{\rm trans}\sim R_s$, and this explains the preponderence of $\ss$
%values of the order of unity.


\end{abstract}
\section{Introduction}

The Crab nebula is certainly the best studied and possibly the
most interesting of the supernova remnants. This is due to the 
fact that it has been detected at an extremely broad range of 
energies, from the radio to the TeV regime. What is of additional
interest is that its expansion is powered by the energy input from
the rapidly rotating pulsar PSR 0531+21 located at its center. As 
such it has been a laboratory for testing our models of pulsars, 
MHD winds, supernova remnants and radiation emission processes.
The Crab is not unique in containing a pulsar at the 
center of a supernova remnant. However, it is unique in the 
efficiency of converting the power output associated with the 
pulsar spin-down to radiation, which reaches 20\%, and to kinetic
energy of the entire remnant which absorbs the remaining 80\%.

Crucial in this efficient conversion of the pulsar power into
radiation is thought to be the presence of a (reverse) strong shock at
an angular distance of $10^{\prime \prime}$ (corresponding to a
distance $ \simeq 3 \times 10^{17}$ cm) from the location of the
pulsar. This shock randomises the highly relativistic upstream MHD
wind which is produced by the pulsar, thereby causing the wind to
radiate away a major fraction of its available energy. The presence of
this strong shock is predicated on the dominance of the relativistic
MHD wind emanating from the pulsar by particles rather than magnetic
field, i.e. that the magnetization parameter (defined below) at the
shock distance has a value $\ss \ll 1$.

The value of \ss has been estimated in a variety of ways. Kennel
\& Coroniti (1984) (hereafter KC) have computed the detailed
structure of the MHD flow downstream from the shock 
and concluded that matching the nebular expansion velocity at 
the nebular edge, using the low \ss expansion of their solution, 
requires that $\ss \simeq 3 \times 10^{-3}$. De Jager \& Harding
(1992) estimated the value of \ss by fitting the spectrum 
and surface brightness of the nebula, under the assumption 
that the emission above 10 GeV is due to inverse Compton 
scattering of lower frequency photons, which presumably 
represent the synchrotron emission from the same electron 
distribution. Their estimate of \ss and the radial distribution 
of the magnetic field are consistent with those proposed by KC.

However, as shown in Rees \& Gunn (1974; eq. 1), matching the 
expansion velocity $v_{\rm{ex}}$ of the nebula at its edge at
$R = R_N$ is just a statement of conservation of the momentum flux
injected by the pulsar wind through an MHD shock at $R = R_s$,
leading to $R_s/R_N \sim (v_{\rm{ex}}/c)^{1/2}$, independent of 
the value of \ss. KC showed that, if in addition $\ss \ll 1$,
one obtains $v_{\rm{ex}}/c \sim \ss$; however, the latter is 
not a condition necessary for matching the nebular expansion
velocity to that of the MHD wind at $R_s$. 

One is therefore led to the conclusion that the small value
of \ss associated with the Crab remnant is not a generic 
property of all remnants of similar morphology but specific
to the Crab. Indeed, the Vela pulsar, located near the 
center of the Vela supernova remnant, has properties not too
different from those of the Crab pulsar (other than its age), 
however, its 
non-thermal nebular emission is a much smaller fraction
of the pulsar spin-down luminosity than it is in the Crab.
The corresponding estimate for \ss in the case of Vela is 
$\ss \simeq 1$, suggesting that prominent (as a fraction 
of the pulsar spin-down) nebular emission is generally associated 
with small values of \ss which allow for the possibility of 
a strong MHD shock in the pulsar wind. 

The values of \ss inferred for the Vela and (even more for) the
Crab pulsar
MHD winds raise the following problem for these winds: the value 
of this parameter near the pulsar light cylinder is estimated to 
be quite high $\s \sim 10^{4-5}$ (Coroniti 1990 and 
references therein). Given that in a MHD wind $B_{\phi} \propto 
1/R$ it is thought that both the magnetic and ram pressures should 
decrease like $1/R^2$, with their ratio thus remaining roughly constant
at the value it attains near the light cylinder. Therefore, values 
$\ss \sim 1$ (let alone $\ss \sim 10^{-3}$) are hard to  understand 
and yet overwhelmingly favored by observation. 

A possible way out of this conundrum is to assume that the 
inertial component of the MHD wind is due to ions rather than
leptons (electrons -- positrons), leading to much smaller values 
of \s even near the light cylinder (Ruderman 1981; Arons 1983).
However, one would then have to find a way of converting 
$\sim 20\%$ of the relativistic proton energy into relativistic
electrons at the MHD shock. 

This problem led  to the suggestion that annihilation of magnetic 
field energy and conversion of the resulting energy into that of the
outflowing particles could indeed provide for the required reduction
in \ss with distance (Coroniti 1990; Michel 1994). 
%The field annihilation
%was actually considered to be inevitable because in a quasispherical
%wind the decrease in the charge density is faster than that of the 
%current density necessary to enforce the ideal MHD condition, thereby
%leading  to field annihilation (Usov 1975?). 
Such a solution is in principle possible (see though Lyubarskii \&
Kirk 2001), however, this process would work only on the magnetic
dipole field component perpendicular to the direction of the pulsar
angular velocity $\Omega$.  The component of the magnetic dipole field
which is parallel to $\Omega$ is simply advected away with no
possibility of such an annihilation. Since the observations
(Aschenbach \& Brinkmann 1975) seem to suggest that, at least for the
Crab, the magnetic dipole is closely aligned with the pulsar rotation
axis, it appears unlikely that a large fraction of the available
magnetic energy could in fact annhilate.

However,  in MHD flows, issues such as the asymptotic (or more 
generally
the position dependent) value of \s are coupled to the global geometry
of the flow. As shown by Heyvaerts \& Norman (1989) for the
non-relativistic
case and by Chieueh , Li \& Begelman (1991; hereafter CLB91)(see also 
Eichler 1993) for the 
relativistic one, these flows tend to asymptotically collimate; the 
associated divergence of lines from conical geometry could then also 
affect the corresponding value of \s through the ``magnetic nozzling"
which would convert magnetic energy to directed motion. It has been
argued in the above references though, that because these flows 
collimate logarithmically in $R$ such a ``nozzling" is not 
observationally
relevant for any plausible astrophysical situation. We also
argue below, that the global geometry of the flow is indeed of 
consequence
for the asymptotic value of \s, however our conclusions differ 
from those presently in the literature.

More recently, Chieueh, Li \& Begelman (1998; hereafter CLB98)
by an asymptotic analysis of the conservation and the 
perpendicular force balance (Grad-Safranov) equations, have argued
for the implausibility of the transition of flows from  high \s 
to low \s under axisymmetric, steady state conditions. To this end
they considered a variety of plausible field geometries and argued
for each of them that divergence of the magnetic field lines necessary 
to achieve a transition from $\s \gg 1$ to $\s \ll 1$ was incompatible
with the balance of the corresponding pressures. While we believe 
their arguments to be sound we also think, as we argue later, that
can also be circumvented.

Motivated by our recent exact solution of the axisymmetric pulsar
magnetosphere (Contopoulos, Kazanas \& Fendt 1999; hereafter CKF), we
have decided to take a closer look at the problem of the entire MHD
wind and its impact on the nebular morphology and dynamics. The
solution of CKF provides the complete, global, magnetic field and
associated electric current structure for an aligned rotator
(Goldreich \& Julian 1969) in the force free (i.e. with negligible
inertia) MHD approximation, including their distribution across the
crucial light cylinder surface.  The main results of that paper are
summarized below:
\begin{enumerate}
\item The magnetosphere consists of a region of closed field lines 
(dipole-like) extending up to the light cylinder, and a region of open
field lines which cross the light cylinder and asymptote to a
monopole-like geometry.
\item The magnetic field structure is continuous and smooth through
the light cylinder, and thus, one cannot anymore invoque `dissipation
zones' at, or around, the light cylinder.  In other words, one has to
look elsewhere for the conversion of magnetic to particle energy and 
by consequence  to the observed high energy radiation in
pulsars.
\item A large scale electric current flows through the
magnetosphere. The current distribution is uniquely determined by (a)
the boundary condition at the origin (in our case a magnetic dipole),
and (b) the requirement of no singularities at the light cylinder. The
large scale electric circuit closes in an equatorial current sheet
which connects to the edge of the polar cap\footnote{Note that this
implies a discontinuity of the toroidal magnetic field component
across the current sheet, which further leads to a discontinuity of
the poloidal field component at the boundary of the closed field line
region (`dead zone'). This latter discontinuity might lead to magnetic
field structure readjustments (as in the magnetar models of Duncan \&
Thompson~1992).}.
\item Outside the light cylinder, the solution with a dipole at the 
origin does not differ much from the well known monopole solution of 
Michel~(1991).
\item Finally, when we numerically checked whether the magnetic field
structure obtained is capable to accelerate the flow of electrons and
positrons from the polar cap, we obtained no significant acceleration.
This last point merits special attention and will be revised, since,
as we said, the observations suggest the presence of a
hyper-relativistic wind of electrons and positrons at large distances.
\end{enumerate}

In \S~2 we outline in detail the so-called \s-problem using
dimensional analysis of the corresponding MHD flow and indicate the
arguments which could lead to its possible resolution. In \S~3, using
the MHD integrals of motion for the B-field geometry associated with
the axisymmtric pulsar magnetosphere, we indicate the evolution of
$\s$ with radius and the eventual values it attains, providing an
explicit resolution to the issue of its magnitude. Finally, in \S~4
we consider other pulsar nebulae for which the value of
\ss has been estimated and our conclusions are drawn.

\section{The $\sigma$ problem}

Since the Crab pulsar is believed to be an almost aligned rotator
(Aschenbach \& Brinkmann 1974), we will adopt the approximation of
axisymmetry in our present discussion.  Below, we provide a summary of
our knowledge of axisymmetric pulsar magnetospheres based on CKF, 
using the Crab pulsar values as fiducial figures.

The field lines that cross the light cylinder emanate from a region
near the pole, the polar cap, and are necessarily open. We calculate
the polar cap radius\footnote{As is obtained in CKF, $\Psi_{\rm open}
=1.36 \Psi_{\rm pc}$, where $\Psi_{\rm open}$ and $\Psi_{\rm pc}$ are
defined as the amount of magnetic flux crossing the distance to the
light cylinder in the relativistic and nonrelativistic
(i.e. undistorted) dipole solution respectively.} to be equal to
\begin{equation}
R_{pc}= \sqrt{1.36}\ r_*\left(\frac{r_*}{R_{lc}}\right)^{1/2}=
0.9\left(\frac{P}{33\ {\rm ms}}\right)^{-1/2}\ {\rm km}
\end{equation}
Here, $r_*=10~{\rm km}$ is the canonical radius of a neutron star, and 
\begin{equation}
R_{lc}=\frac{cP}{2\pi}=
1576\left(\frac{P}{33\ {\rm ms}}\right)\ {\rm km}
\end{equation}
is the light cylinder radius ($P$ the period of the neutron star
rotation)\footnote{Henceforth, we will denote cylindrical radii with
capital $R$, and spherical radii with small $r$.}. Obviously,
$R_{pc}\ll r_*\ll R_{lc}$. At the footpoints of the magnetic
field lines on the polar cap, the magnitude of the magnetic field $B_*$
is of the order of $10^{12}$~G, the number density $n_*$ of
electrons/positrons in the outlfowing wind is equal to
\begin{equation}
n_* = \kappa n_{GJ}\equiv \kappa \frac{B_*}{ePc}= 
2\times 10^{16}\left(\frac{\kappa}{10^4}\right)
\left(\frac{B_*}{10^{12}\ {\rm G}}\right)
\left(\frac{P}{33\ {\rm ms}}\right)^{-1}
{\rm cm}^{-3}
\ ,
\end{equation}
and their Lorentz factor $\gamma_*$ is of the order of 200 (see
below).  Here, $e$ is the electron charge. The multiplicity
coefficient $\kappa$ expresses how many times the wind density
surpasses the so called Goldreich-Julian density $n_{GJ}$ at the base
of the wind. The physics that determine $\kappa$ and $\gamma_*$ lie
outside the context of ideal magnetohydrodynamics (CKF). In what
follows, values of $\kappa \sim 10^{3-4}$ and $\gamma_* \sim 200$
are adopted from the cascade
models of Daugherty \& Harding (1982) who followed in detail cascades
of high energy electrons in pulsar magnetospheres.

The open field lines contain an amount of magnetic flux
\begin{equation}
\Psi_{\rm open}=\pi R_{pc}^2 B_*
=2.7\times 10^{12}\left(\frac{P}{33\ {\rm ms}}\right)^{-1}
\left(\frac{B_*}{10^{12}\ {\rm G}}\right) {\rm G}\ {\rm km}^2
\label{Psipc}
\end{equation}
and an electron/positron wind with mass loss rate
\begin{equation}
\dot{M}=\pi R_{pc}^2 n_* m_e
=7.8\times 10^{-31} \kappa
\left(\frac{P}{33\ {\rm ms}}\right)^{-2}
\left(\frac{B_*}{10^{12}\ {\rm G}}\right) {\rm M_{\odot}}\ {\rm 
yr}^{-1}
\label{Mdot}
\end{equation}
from each polar cap ($m_e$ is the electron rest mass). We have assumed
here an almost uniform `loading' of the polar cap field lines with
matter.  The wind carries a kinetic energy flux
\begin{equation}
W_{\rm Kinetic}
= \gamma \dot{M} c^3
= 7\times 10^{-5} \kappa \left(\frac{\gamma}{200}\right)
\left(\frac{P}{33\ {\rm ms}}\right)^{-2}
\left(\frac{B_*}{10^{12}\ {\rm G}}\right) L_\odot
\ ,
\label{KE}
\end{equation}
and the magnetic field carries a Poynting flux
\[
W_{\rm Poynting} = \frac{\Omega}{2\pi c}\int_0^{\Psi_{\rm open}}
I(\Psi){\rm d}\Psi
= \frac{\Omega}{2\pi c}fI\Psi_{\rm open}
\]
\begin{equation}
=10^4 f\left(\frac{I}{I_*}\right)
\left(\frac{P}{33\ {\rm ms}}\right)^{-4}\left(\frac{B_*}{10^{12}\ {\rm 
G}}
\right)^2L_\odot
\label{PF}
\end{equation}
from each polar cap (Okamoto 1974). Here,
\begin{equation}
I_*\equiv \frac{\Omega\Psi_{\rm open}}{2}\equiv 
\frac{1}{4}\cdot  en_{GJ}c\cdot \pi R_{pc}^2
\end{equation}
and $I$ are the total amount of electric current flowing through the
polar cap and the magnetosphere respectively.  As we will see,
contrary to $\dot{M}$ and $\Psi_{\rm open}$, $I$ {\em cannot be a
conserved quantity along the wind}.  $f$ is a factor of order unity
which depends on the distribution of the electric current $I(\Psi)$
across open field lines ($f=0.67$ for the exact monopole solution of
Michel~(1991), and the numerical solution of CKF).  The energy 
reservoir
is obviously the neutron star spindown energy loss rate
\begin{equation}
W_{\rm Spindown}=W_{\rm Kinetic*}+W_{\rm Poynting*}=W_{\rm
Kinetic}+W_{\rm Poynting}
\label{SD}
\end{equation}
at all distances. The magnetization parameter $\sigma$ is thus defined
as
\begin{equation}
\sigma \equiv \frac{W_{\rm Poynting}}{W_{\rm Kinetic}}=
\frac{\Psi_{\rm open} I}{Pc \gamma \dot{M} c^3}=
\left(\frac{I}{I_*}\right)
\left(\frac{\gamma_*}{\gamma}\right) \sigma_*\ .
\label{sigma}
\end{equation}
Here,
\begin{equation}
\sigma_* \equiv
\frac{W_{\rm Poynting *}}{W_{\rm Kinetic *}} =
\frac{efI_*}{\kappa \gamma_* m_e c^3} =
\frac{1.6\times 10^8}{\kappa}f
\left(\frac{P}{33 {\rm ms}}\right)^{-2}
\end{equation}
is the value of the magnetization parameter near the surface of the
neutron star.

As we discussed in the introduction, at a distance 
\begin{equation}
r_s \sim 10^9 R_{lc}\ ,
\end{equation}
the pulsar wind slows down in a (reverse) shock, and approximately
20\% of the neutron star spin-down luminosity is converted into
optical to $\gamma$-ray radiation. The infered value of the
magnetization parameter is
\begin{equation}
\sigma_s \sim 3\times 10^{-3}\ .
\end{equation}

Using eq.~(\ref{SD}) divided through with $W_{\rm Kinetic *}$
\begin{equation}
\sigma_* + 1 = \frac{\gamma}{\gamma_*}(\sigma + 1)\ ,
\label{sigmas}
\end{equation} 
And making use of eq.~(\ref{sigma}), it is straightforward to see that
the dramatic decrease in $\sigma$ observed in the Crab could reasonably
take place (based on whether the associated current is conserved or 
not)
% must take place *** yianni, one can in principle have a discontinuous
% transition as discussed by clb98
in
two distinct regimes:
\begin{enumerate}
\item $\sigma$ decreases from $\sigma_*\sim 10^{4}$ to $\sigma_{\rm
trans}=1$.  In that regime, the magnetospheric electric current is
almost conserved, i.e.
\begin{equation}
\frac{I_{\rm trans}}{I_*}= \frac{1}{2}\ ,
\end{equation}
whereas the wind Lorentz factor increases by a factor
\begin{equation}
\frac{\gamma_{\rm trans}}{\gamma_*}= \frac{\sigma_*}{2}\ ,\ {\rm i.e.}\ 
\gamma_{\rm trans}=3\times 10^{6}\ .
\end{equation}
\item $\sigma$ decreases from $\sigma_{\rm trans}=1$ to $\sigma_s
=3\times 10^{-3}$. 
In that regime, the magnetospheric electric current decreases by 
a factor
\begin{equation}
\frac{I_s}{I_{\rm trans}}= \frac{2}{\sigma_*}\ ,
\end{equation}
whereas the wind Lorentz factor remains almost constant, i.e.
\begin{equation}
\frac{\gamma_s}{\gamma_{\rm trans}}= 2\ ,\ {\rm i.e.}\ 
\gamma_s=6\times 10^{6}\ .
\end{equation}
\end{enumerate}
In fact, detailed modelling of the nebula's spectrum yields a shock
upstream wind Lorentz factor $\gamma_s \sim 3\times 10^6$, in agreement
with the above (De Jager \& Harding~1992).

Unfortunately, our theoretical understanding is not up to date with
the above observational facts. The problem is that there is no
indication for growth in the Lorentz factor in the inner
magnetosphere, and the electric current is a conserved quantity in a
force-free magnetosphere.  Acceleration models invoquing dissipation
zones near the light cylinder are not convincing anymore after CKF
(e.g. 
%Mestel {\em et al.} 1985, 
Beskin, Gurevich \& Istomin 1993), and models showing MHD acceleration
at large scales define a-priori the field geometry (e.g. Takahashi \&
Shibata~1998).  There exist also MHD models showing no acceleration at
large scales, but this might have to do with their numerical
extrapolation from small to large scales (e.g. Bogovalov \&
Tsinganos~1999, Bogovalov~2001). Our understanding is that current MHD
acceleration models are incomplete.

Is it possible to account for the large scale acceleration of the
pulsar wind in the context of ideal axisymmetric special-relativistic
steady-state magnetohydrodynamics? We believe that the answer is yes,
so let us now study the basic equations of the problem, focusing our
analysis on the energy flux conservation equation along open field
lines. As we will see, in order to reveal the flow acceleration,
one has to be particularly careful when one takes limits of that
equation at large distances.

\section{The linear accelerator}

Energy flux conservation
implies that
\begin{equation}
\gamma \left(1-\frac{R}{R_{lc}}\frac{v_\phi}{c}\right)=
\gamma_*
\label{Bernoulli}
\end{equation}
along any open field line (e.g. Okamoto 1978, Mestel \& Shibata 1994,
Contopoulos 1995), with $\gamma_*$ the initial value of the electron
Lorentz factor ($\gamma_* \sim 200$ as discussed above). 
This is just the differential form of the energy flux conservation 
equation~(\ref{SD}).  The induction equation further gives that
\begin{equation}
\frac{v_\phi}{c}=
\frac{R}{R_{lc}}+\frac{v_p}{c}\frac{B_\phi}{B_p}\ .
\label{vphi}
\end{equation}
In order to simplify the notation, we will concentrate our discussion
on the last open field line along the equator.  

As we argued above, in order to determine the evolution of the flow
Lorentz factor $\gamma$ with distance, 
it is reasonable to consider the two different distance regimes 
in the large scale pulsar magnetosphere we determined above 
(force-free,
non force-free),
and make different approximations in each of them.  Let us first
consider distances much larger than the light cylinder, where $\sigma
\sgreat 1$ (and most likely $\s \gg 1$).
We show in the Appendix that, under force--free conditions
(i.e. negligible inertia),
\begin{equation}
B_\phi = -\frac{R}{R_{lc}}B_p
\label{as}
\end{equation}
when $R\gg R_{lc}$ (e.g. Okamoto 1997).  This is identically valid in
the analytical monopole solution (Michel~1991). It is also identically
valid in the asymptotic monopole-like part of the more realistic
solution with a dipole at the origin (CKF).  Eqs.~(\ref{Bernoulli})
and (\ref{vphi}) then yield
\[
\gamma\left[1-\left(\frac{R}{R_{lc}}\right)^2\left(1-
\frac{v_p}{c}\right)\right]=\gamma_*\ ,
\]
which further yields
\begin{equation}
\gamma=\left[\gamma_*^2+\left(\frac{R}{R_{lc}}\right)^2\right]^{1/2}
\rightarrow \frac{R}{R_{lc}}
\label{lineargamma}
\end{equation}
for $R\gg R_{lc}$. This is a very important result. In addition to 
providing for the radial dependence of the wind's Lorentz factor,  
it also makes clear why CKF came (erroneously) to the conclusion 
that there is no acceleration across the light cylinder: as long as 
$R/R_{lc}<\gamma_*=200$ (the value used by CKF based on the results
of pair cascades in pulsars), the associated growth in $\gamma$ is 
imperceptible (they would have found the increase had they chosen
say $\gamma_* \sim 1$).  The new result here is {\em the linear growth 
of the Lorentz factor $\gamma$ with distance $R$}, for
$R/R_{lc}\gg\gamma_*$. The reader can check that the pure monopole
solution (Michel~1991) also shows this effect! 
%*** text addition. i commented out the old text; here is the new one.
%This dependence indicates the gradual dominance of inertial effects
%while the flow is essentially force-free. Therefore, for an eventual
%conversion of the flow from magnetic to inertial dominance 
%no extraordinary gradients are necessary. All terms in the 
%concervation and cross field balance equations are of similar 
%magnitude, thereby circumventing the arguments put forward in 
%CLB98 (the LHS of their Eq. (22) should also contain  mass flux
%terms of not too different magnitude). 
This dependence indicates that the conversion of the flow energy
from magnetic to kinetic is gradual. In fact, the conversion 
almost to equipartition takes place while the flow is essentially
force-free and therefore our solution can be trusted. The main
difference of our analysis with that of CLB98 is that upstream 
of their transition region the flow is essentially non-relativistic
while ours has already a Lorentz factor $\gamma \sgreat 10^5$.
This constitutes the pivotal point in circuvmenting the 
analysis of CLB98, who argued against a transition from magnetic
to inertial dominance of the flow at large distances.

The linear growth, however, cannot continue beyond a distance
\begin{equation}
\gamma_* \sigma_* R_{lc}
=2R_{\rm trans}=3\times 10^6 R_{lc}\ll r_s\ ,
\end{equation}
at which the Lorentz factor reaches the asymptotic value implied by
mass conservation and the observed spin-down luminosity.
The problem we are presented with has arisen from our neglect of
matter in our assumption of negligible inertia (i.e. force--free)
conditions. Note that eq.~(\ref{as}) is not valid at the distance
where inertial and magnetic forces become comparable\footnote{As we 
will
see, eq.~(\ref{as}) becomes again asymptotically valid at the distances 
where inertial effects dominate. It is important to emphasize here that 
if
we use eq.~(\ref{as}) in conjunction with eq.~(\ref{Bernoulli}) in
that latter regime, we will lose the effect of inertia, and will thus
be led to the erroneous conclusion of unlimited linear growth in 
$\gamma$.}.
We must therefore proceed with caution through the energy conservation
equation written as eq.~(\ref{sigmas}).  In the case of the Crab
pulsar, we obtain
\begin{equation}
\frac{\sigma_s}{\sigma_{\rm trans}}=
\sigma_s = \frac{I_s}{I_{\rm trans}}\frac{\gamma_{\rm trans}}
{\gamma_s}= \frac{1}{2}\frac{I_s}{I_{\rm trans}}
= \frac{1}{2}\frac{(R B_{\phi})|_s}{(R B_{\phi})|_{\rm trans}}
= 3\times 10^{-3}\ 
\label{collimation}
\end{equation}
(see also Okamoto~1997 and references therein). The reader can check
that the same result can also be obtained through the differential
form of the Bernoulli equation [eq.~\ref{Bernoulli}]).

What does eq.~(\ref{collimation}) imply for the wind morphology at
those distances? It is shown in the Appendix that, when $\sigma\ll 1$,
$B_\phi\rightarrow -(R/R_{lc})B_p$, and thus
\begin{equation}
\frac{(R^2 B_p)|_s}{(R^2 B_p)|_{\rm trans}}
\sim 6\times 10^{-3}\ .
\label{collimation2}
\end{equation}
In other words, beyond $R_{\rm trans}\sim 5\times 10^5\ R_{lc}$, $B_p
R^2$ does not remain constant but decreases with distance, and
consequently, {\em field/flowlines should diverge away from monopolar 
geometry towards the axis of symmetry}. $R$ in eq.(\ref{collimation2}) 
is the cylindrical radius, so assuming that $B_p$ evolves roughly as 
$1/r^2$, this implies that, along a field/flowline,
\begin{equation}
(R/r)|_s \approx 10\% \cdot  (R/r)|_{\rm trans}\ ,
\end{equation}
or equivalently, a collimation by a factor of 10 in the
cylindrical radius (the cylindrical radius at the
shock spherical radius $r_s$ will be about 10 times smaller
than what it would be if the field/flow lines continued
in the radial spherical direction from the transition
distance).

The precise field geometry can only be determined through a full
solution of the Grad-Shafranov equation, whose solution we defer
to a future publication.  The degree of collimation implied by
eq.~(\ref{collimation2}) is not unreasonable, if one considers the 
Hubble
Space Telescope (HST) observations of the Crab nebula in the optical,
and the Chandra observations in X-rays. One can clearly see there the
presence of a collimated polar flow, within a distance from the axis
of symmetry of the order of ten times smaller than the overall size
of the Crab nebula. %*** (pls check the next change same paragraph)
Finally, for a much more detailed matching of the theoretical
results to observation one should also consider the effects of
the remnant into which the MHD wind is plowing. It is not obvious 
to the authors that every detail can be accounted for in terms of 
MHD winds and their self-collimation while ignoring the effects
of the outlying medium.

Let us summarize our results for the Crab pulsar wind.  The wind
Lorentz factor grows linearly with distance up to a distance of the
order of $10^6\cdot R_{lc}$ where $\sigma \sim 1$.  The wind/field
geometry remains almost monopolar up to that distance. Beyond that,
the wind collimates drastically towards the direction of the axis of
symmetry.  Its Lorentz factor remains close to its asymptotic value,
and $\sigma$ reaches its inferred value of $3\times 10^{-3}$ at the
shock distance $10^9 R_{lc}$.  Our present conclusion, namely the
inevitable convergence of field/flowlines towards the axis of symmetry
in order for the flow to accelerate to $\s\ll 1$, is not original (see
Okamoto~1997 and references therein).  Nevertheless, we are now in a
position to make definite predictions about the degree of field/flow
collimation, without solving the full non-force-free problem.

\section{Conclusions, Discussion}

We have presented above the spatial evolution of the kinetic
energy associated with the wind from an axisymmetric pulsar 
magnetosphere. By solving the energy equation in the regions
in which our exact solution of the MHD equations (CKF), based
on the force-free assumption, is valid we indicated how the gradual 
acceleration of the expanding wind can lead to an equipartition between 
the magnetic and particle fluxes thus effecting the efficient 
conversion 
of magnetic to particle energy. This provides a first (to our 
knowledge)
concrete example which exhibits such a conversion from Poynting to 
particle energy flux. We further indicated that at distances 
larger than that at which equipatition is established, the inertial
effects should become important leading to a collimation of the 
wind and a further decrease in the ratio of magnetic to particle
fluxes in agreement with HST/Chandra/CGRO observations of the 
Crab pulsar/nebula. Our work thus provides a straightforward resolution 
of the long standing $\sigma-$problem namely that of the particle
over the magnetic flux dominance of this object. 

As discussed by Mestel (1999), the issue of the precise magnetospheric
geometry and the associated evolution of the electron Lorentz factor
is a coupled problem; he then raises the question of whether there
may indeed be solutions in which the wind achieves its asymptotic
value  close to the light cylinder with the entire solution collimating
at that distance but at the cost of requiring a domain in which 
dissipation enforces  a local departure from the perfect conductivity
condition. The work of CKF and our present argumentation indicates 
that this is not necessary and it is, in addition, compatible with 
the observations which require very little emission at radii smaller
than $r_s$. 

While the entire magnetosperic solution (including
the inertial and external medium effects) is desirable in order to 
assess their mutual interactions and their effects on the precise
geometry and flow dynamics, we believe that near (i.e. at $R \sles
10^4 R_{lc}$) the pulsar the flow geometry is so strongly dominated
by its presence and the magnetic field, that the solution 
of CKF is essentially correct. As we argued earlier, we believe
that important as the analysis of CLB98 is,  its arguments  
could be circumvented simply because at the transition region the flow
is almost in equipartition rather than magnetically dominated,
as it is usually considered when analysing this situation (prior
to our work the genenal assumption was that for a conical flow
$\rho \gamma, B^2_{\phi} \propto R^{-2}, ~B^2_{\phi} \gg \rho \gamma$; 
efficient acceleration then required a very tightly wound-up $B^2_{\phi}$
which would convert to kinetic density in a short distance; this is not
any more necessary). 

The results presented above are fairly general and could be applied to 
the case of other less well studied pulsar magnetospheres. Important 
parameters of our problem which are  expected vary from pulsar to 
pulsar are the period $P$, the multiplicity coefficient $\kappa$, 
the initial flow Lorentz factor $\gamma_*$. Knowing $P, \kappa, 
\gamma_*$, 
one can estimate $\s_*$, and from it the characteristic distance 
\begin{equation}
R_{\rm trans} \sim \gamma_* \s_* R_{lc} \propto R_{lc}/ \kappa
\end{equation}
of the problem. Fortunately, not all these parameters are indepedent.
The Lorentz factor of the pair resulting from the magnetospheric 
cascade $\gamma_*$ as well as the multiplicity $\kappa$ depend
on the magnetic field $B$ and the pulsar period $P$ (Zhang \& Harding
2000), with the multiplicity generally decreasing with decreasing 
$B$ and increasing $P$, implying that there is in reality much less 
freedom in the problem. It would be of interest to compare the 
expected values of $R_{\rm trans}$ with the distance to the synchrotron 
nebula reverse shock $R_s$ obtained from observations in other 
such nebulae. If $R_{\rm trans}$ is found to be $\ll R_s$, then
our previous analysis applies, the flow should, like in the Crab, 
collimate
towards the axis of symmetry, and at the same time $\s$ should decrease
to values $\ll 1$.

However, the Crab remnant seems to be a singular example, in that 
the neutron star has remained very close to the center of the 
associated
supernova remnant, and the synchrotron nebula has had enough time to
grow to substantial distances. Furthermore, the entire nebula is 
powered
exclusively by the pulsar, a fact not necessarily true with other 
remnants. The rather large peculiar velocities of these pulsars 
lead to geometries which are significantly affected by the pulsar 
motion
and make similar comparisons difficult. The closest other remnant is 
that associated with the Vela pulsar for which the value of the 
magnetization parameter $\sigma$ was estimated from detailed spectral
fitting to be $\sigma \sim 1$ (de Jager et al. 1996). Similar 
conclusion was reached by Helfand et al. (2001) who analyzed the 
spatially resolved Chandra images of this source. It may therefore 
be that the low value of $\sigma$ associated with the Crab nebula
is specific to the conditions prevailing in this remnant.

The specific radial dependence of the pulsar wind's Lorentz factor
is expected to have additional observational consequences concerning
the emission of high energy radiation in systems containing pulsar 
winds. For example, Bogovalov \& Aharonian (2000) computed the 
upComptonization of soft photons to TeV energies in the Crab through
their interaction with the expanding MHD wind while Tavani \& Arons
(1997) and Ball \& Kirk (2000) computed the corresponding radiation 
expected by the radio-pulsar Be star binary system PSR B1259-63
through the interaction of the relativistic wind with the photon 
field of the companion. We expect  these predictions to be modified 
considerably in view of our present results. For example the 
above works assume that the wind achieves its asymptotic Lorentz
factor shortly beyond the light cylinder, with the IC luminosity
given by $L_{IC} \propto \tau(\gamma) L_s$ where $L_s$ is the 
soft photon luminosity and $\tau(\gamma)$ is the optical depth
for scattering by electrons of Lorentz factor $\gamma$. Clearly,
our proposed linear evolution of the wind Lorentz factor $\gamma$
would lead to a very different dependence for the optical depth
of electrons with a given Lorentz factor in the expanding wind.
This in turn should lead to a high energy gamma ray spectrum 
of very different form than that obtained under the assumption
of constant electron Lorentz factor used in the aforementioned works.
We expect that careful modeling of the resulting spectra and comparison 
to (future) high energy $\gamma-$ray observations will allow to
confirm or disprove the proposed linear with $r$ evolution of the 
wind Lorentz factor.

We would like to acknowledge useful discussions with Alice Harding 
and Okkie de Jager.

%{\bf Yianni i have moved this part here rather than leaving it 
%out, until the final version}

%. For A result of this
%translational motion are the complicated and smaller sizes of their 
%associated synchrotron nebulae, which also imply small values for 
%the reverse shock distance.  Such an example is the Vela remnant and 
%pulsar, 
%for which one obtains $R_{lc}=10^9\ {\rm cm}$, $\gamma_*=200$, 
%$\s_*=10^4$ 
%(Harding~1981), and $R_{\rm trans}\sim \gamma_* \s_* R_{lc}\sim 
%10^{15}\ 
%{\rm cm}$, comparable to $R_s\sim
%10^{16}\ {\rm cm}$({\bf reference from X-ray observations}).  
%In this case the high pulsar velocity makes the magnetospheric 
%topology much more complicated with a bow shock apparent in the
%X-ray image due to the pulsar motion through the ambient gas 
%of the SN remnant.
%We speculate that this has prevented the collimation of the flow
%and the associated reduction of $\s$ beyond the value $\sim 1$ as
%it happened with the Crab.

%T
%In that
%case, the flow did not have time to collimate significantly beyond
%$R_{\rm trans}$, and its associated $\s$ did not have the chance to
%decrease significantly below unity. This simple argument accounts
%naturally for the decrease of $\s$ to values of the order of unity in
%most pulsars, with Crab being a singular exception.

\newpage
\section*{Appendix}

We derive here simple relations between the toroidal and poloidal
magnetic field components at large distances from the light cylinder.

We will first consider the regime where force-free conditions are
valid ($\s\geq 1$), and according to CKF, the field has attained a
monopolar distribution, where all physical quantities become functions 
of
the spherical angle $\theta$. In that regime, the pulsar equation takes
the simple form (eq.~15 in CKF)
\begin{equation}
\frac{{\rm d}^2\Psi}{{\rm d}t^2}=-\frac{{\rm d}\Psi}{{\rm d}t}
\frac{1+2t^2}{t(1+t^2)}+\frac{4R_{lc}^2I\frac{{\rm d}I}{{\rm 
d}\Psi}}{c^2
t^2(1+t^2)}\ ,
\label{asymptote}
\end{equation} 
where, $t\equiv\tan\theta$, with boundary conditions $\Psi(t=0)=0$,
and $\Psi(t=\infty)=\Psi_{\rm open}$. Bearing in mind that
$I(\Psi=0)=0$ (i.e. no singular current along the axis of symmetry),
eq.~(\ref{asymptote}) can be integrated to yield
\begin{equation}
I=-\frac{c}{2 R_{lc}}\frac{{\rm d}\Psi}{{\rm d}\theta}\sin\theta
\label{Isolution}
\end{equation}
(the reader can check that $I$ and $B_p$ point in opposite directions,
thus the minus sign).  It is now straightforward to see that
\begin{equation}
\frac{B_\phi}{B_p}\equiv \frac{\frac{cI}{2R}}{-\frac{1}{r^2 \sin\theta}
\frac{{\rm d}\Psi}{{\rm d}\theta}}=-\frac{R}{R_{lc}}\ .
\label{ratio}
\end{equation}
This is a very general result, which does not depend on whether we
have a dipole or a (split) monopole at the center (the reader can
check that eq.~(\ref{ratio}) is directly satisfied for the Michel~1991
split monopole solution).

We will next consider the symptotic regime where force-free conditions
are not valid anymore ($\s\ll 1$, or $4\pi\rho \gamma v_p^2/B_p^2 \gg
1$, where $\rho$ is the matter density in the observer's fixed
frame). In that regime, we need to keep all the terms in the
expression for $B_\phi$ (e.g. Contopoulos~1994)
\begin{equation}
B_\phi=\frac{\frac{cI}{2R}-\frac{4\pi \rho \gamma v_p}{B_p}
R\Omega}{1-\frac{4\pi \rho \gamma v_p^2}{B_p^2}}\rightarrow
-\frac{\frac{cI}{2R}-\frac{4\pi \rho \gamma c}{B_p}
R\Omega}{\frac{4\pi \rho \gamma c^2}{B_p^2}}\rightarrow
\frac{R}{R_{lc}}B_p\ .
\label{ratio2}
\end{equation}
The reader should keep in mind that, in order to obtain
eq.~(\ref{ratio2}), we have taken a limit, which is not the case for
eq.~(\ref{ratio}).

\end{document}